\documentclass[conference]{IEEEtran}
\IEEEoverridecommandlockouts
\usepackage{cite}
\usepackage{amsmath,amssymb,amsfonts}
\usepackage{algorithmic}
\usepackage{graphicx}
\usepackage{textcomp}
\usepackage{xcolor}
\usepackage{CJKutf8}  
\usepackage{tabularx}
\usepackage{booktabs}
\usepackage{ragged2e}  
\usepackage{microtype}  
\usepackage{array}     
\usepackage{hyperref}
\usepackage{graphicx}
\usepackage{subcaption}
\usepackage{listings}
\usepackage{xcolor}
\usepackage{booktabs}  
\usepackage{array}     
\usepackage{enumitem}  
\def\BibTeX{{\rm B\kern-.05em{\sc i\kern-.025em b}\kern-.08em
    T\kern-.1667em\lower.7ex\hbox{E}\kern-.125emX}}
\begin{document}
\begin{CJK}{UTF8}{gbsn}  

\title{TriFusion-LLM: Prior-Guided Multimodal Fusion with LLM Arbitration for Fine-grained Code Clone Detection\\
}

\author{
\IEEEauthorblockN{1\textsuperscript{st} Mengdi Li}
\IEEEauthorblockA{\textit{School of Computer Science and Technology} \\
\textit{University of Chinese Academy of Sciences}\\
Beijing, China \\
limengdi24@mails.ucas.ac.cn}

\and
\IEEEauthorblockN{2\textsuperscript{nd} Yuming Liu}
\IEEEauthorblockA{\textit{School of Cyber Engineering} \\
\textit{Xidian University}\\
Xi’an, China \\
24009290010@stu.xidian.edu.cn}

\and
\IEEEauthorblockN{3\textsuperscript{rd} He Wang}
\IEEEauthorblockA{\textit{School of Cyber Engineering} \\
\textit{Xidian University}\\
Xi’an, China \\
hewang@xidian.edu.cn}

\and
\IEEEauthorblockN{4\textsuperscript{th} Zifeng Xu}
\IEEEauthorblockA{\textit{Hainan University} \\
\textit{Hainan University}\\
China \\
zfxu@hainanu.edu.cn}

\and
\IEEEauthorblockN{5\textsuperscript{th} Yuqing Zhang*}
\IEEEauthorblockA{\textit{School of Computer Science and Technology} \\
\textit{University of Chinese Academy of Sciences}\\
Beijing, China \\
zhangyq@nipc.org.cn}
}
\maketitle

\begin{abstract}
Code clone detection (CCD) supports software maintenance, refactoring, and security analysis. Although pre-trained models capture code semantics, most work reduces CCD to binary classification, overlooking the heterogeneity of clone types and the seven fine-grained categories in BigCloneBench. We present Full Model, a multimodal fusion framework that jointly integrates heuristic similarity priors from classical machine learning, structural signals from abstract syntax trees (ASTs), and deep semantic embeddings from CodeBERT into a single predictor. By fusing structural, statistical, and semantic representations, Full Model improves discrimination among fine-grained clone types while keeping inference cost practical. On the seven-class BigCloneBench benchmark, Full Model raises Macro-F1 from 0.695 to 0.875. Ablation studies show that using the primary model's probability distribution as a prior to guide selective arbitration by a large language model (LLM) substantially outperforms blind reclassification; arbitrating only ~0.2\% of high-uncertainty samples yields an additional 0.3 absolute Macro-F1 gain. Overall, Full Model achieves an effective performance-cost trade-off for fine-grained CCD and offers a practical solution for large-scale industrial deployment.
\end{abstract}

\section{Introduction}
Code Clone Detection (CCD), the core technology for identifying functionally equivalent or similar code fragments, is indispensable for diverse software engineering tasks, including software refactoring, bug propagation analysis, and security auditing~\cite{ROY2009470}. Over the past two decades, the detection paradigm has evolved from efficient token-based matching (e.g., CCFinder~\cite{WANG2023111618}) to syntax-based structural alignment via Abstract Syntax Trees (ASTs, e.g., Tamer~\cite{Hu2023FineGrainedCC}), and has more recently converged toward Pre-trained Models (PTMs)\cite{feng-etal-2020-codebert, guo-etal-2022-unixcoder}. This evolutionary trajectory is driving CCD from static pattern matching toward deep, dynamic logical analysis\cite{10576803, liang2025hyclonebridgingllmunderstanding}.
Despite the significant strides made by PTMs and their derivative multi-modal fusion frameworks (e.g., DSFM~\cite{10.1145/3597503.3639215} and FCCA~\cite{9146274}) in capturing sophisticated semantic features, contemporary literature frequently collapses CCD into a simplistic binary classification task. Such oversimplification renders models inadequate when navigating the fine-grained 7-class taxonomy defined by \textit{BigCloneBench} (BCB), where decision boundaries between categories with nearly identical syntactic similarities remain notoriously blurred. This limitation originates from a "capability gap" across two dimensions. First, pure neural fusion models are susceptible to "Semantic Drift"\cite{7332459} during the processing of complex clone pairs; high-level functional embeddings, lacking explicit grounding in programmatic physical facts, often fail to strictly align with the rigid quantitative similarity thresholds of BCB. Second, while recent efforts like Toma\cite{10.1145/3597503.3639114} demonstrate that lexical features and Machine Learning (ML) can serve as stable discriminative anchors, these methods exhibit significant cognitive blind spots when confronted with Type-4 semantic clones involving radical structural overhauls. When literal statistical evidence conflicts with deep semantic intuition at the decision margin, existing architectures lack an adaptive arbitration mechanism to resolve uncertainty and facilitate high-order logical rectification while maintaining retrieval efficiency.
To bridge these gaps, we propose a multi-modal fusion framework that integrates ML-based heuristic priors, AST structural skeletons, and CodeBERT semantic embeddings. Our study is structured around two pivotal research questions:
\begin{itemize}
\item \textbf{RQ1:How effective is the proposed multi-modal fusion framework, and what is the contribution of each component to the overall performance?} 
This question evaluates the framework's global efficacy across the comprehensive 7-class taxonomy (Labels 0--6). By benchmarking against state-of-the-art models, we quantify how lexical, structural, and semantic features synergistically enhance performance metrics, providing a granular attribution analysis across the similarity gradients.
\item \textbf{RQ2:How does the confidence-aware adaptive arbitration mechanism resolve discriminative conflicts?} This question investigates the use of higher-order logical reasoning to reconcile conflicting feature signals. Based on a quantitative analysis of the primary model's probability distributions and confidence intervals, we derive an adaptive triggering policy. Our evaluation focuses on the mechanism's capacity to rectify "gray-zone" samples while exploring the trade-off between detection fidelity and computational overhead.
\end{itemize}

The primary contributions of this work are summarized as follows:
The primary contributions of this work are summarized as follows:
\begin{itemize}
\item \textbf{A Multi-modal Fusion Paradigm:} We architect a composite detection framework that synergizes lexical statistics, structural logic skeletons, and deep attentional semantics. This multi-dimensional integration effectively mitigates the "semantic drift" prevalent in fine-grained clone discrimination.
\item \textbf{Rigorous Empirical Evaluation:} Extensive evaluations on the \textit{BigCloneBench} (BCB) dataset demonstrate the framework's state-of-the-art performance in 7-class classification. We further provide a comprehensive mechanism analysis dissecting the performance gains across diverse clone topologies.
\item \textbf{Adaptive Cognitive Arbitration:} We innovatively introduce an LLM-based arbitrator triggered by uncertainty quantification. By leveraging the primary model's output probability distributions as contextual guidance to rectify low-confidence conflictual instances, \textbf{this informed arbitration significantly outperforms baseline approaches that rely on blind, standalone LLM re-classification.}
\item \textbf{A Data-driven Triggering Strategy:} Guided by validation set statistics, we formulate a quantitative, adaptive triggering criterion that achieves an optimal equilibrium between discriminative fidelity and computational inference overhead.
\item \textbf{Open-source Implementation:} To facilitate reproducibility and further research, the complete implementation and experimental scripts are publicly available at: \url{https://github.com/LMD4790/TriFusion-LLM}.
\end{itemize}
\section{Related Work}
\label{sec:part2}

\subsection{Taxonomy and Benchmarking of Code Clones}
\label{subsec:2_1}
Formal taxonomies underpin clone detection evaluation. While Roy et al.’s four-tier scheme \cite{ROY2009470} is foundational, it lacks threshold clarity between Type-3 and Type-4. BigCloneBench (BCB) \cite{7332459, 7816515} resolves this through stratified similarity gradients (VST3, ST3, MT3, and WT3/T4), becoming the de facto standard for fine-grained discrimination. Despite these advancements, most research collapses clone detection into a simplistic binary task, masking nuanced structural variations. Investigating BCB’s fine-grained failure modes is therefore vital for identifying performance bottlenecks and steering the field toward precise, interpretable paradigms.
\begin{figure*}[!t]
\centering
\includegraphics[width=1\textwidth]{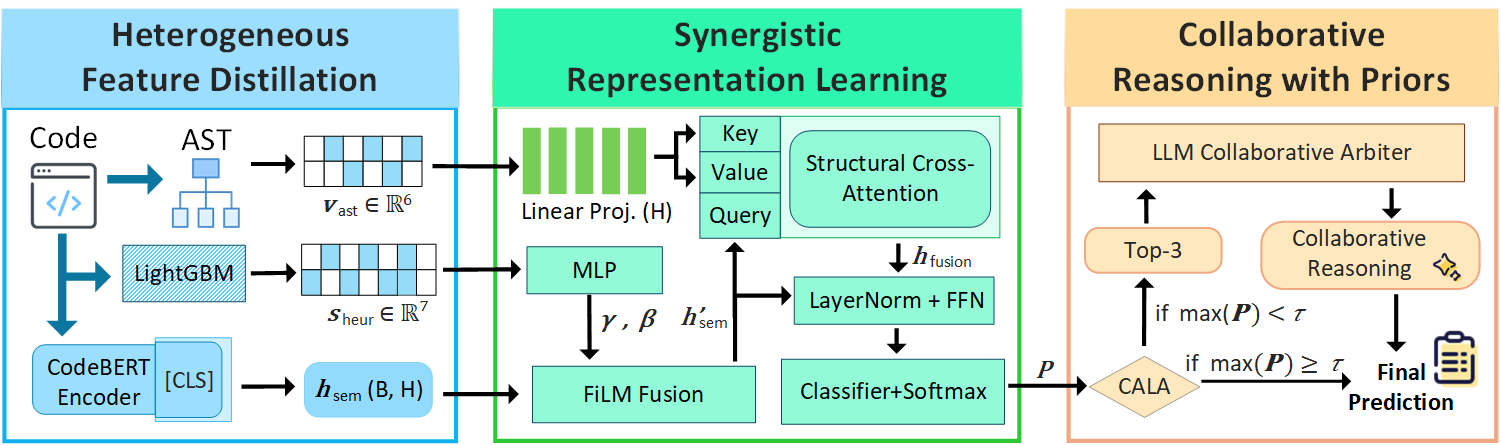}
\caption{Overall pipeline of the proposed multi-dimensional clone detection framework.}
\vspace{0.3em} 
\footnotesize 
\textbf{Note:} CALA=Confidence-aware Logic Arbitration.
\label{fig:all}
\end{figure*}
\subsection{Conventional and Augmented Code Clone Detection}
\label{subsec:2_3}
Extant research in code clone detection emphasizes the integration of lexical rules and structural constraints to foster interpretable identification. CodeGuard~\cite{Glani2024CodeGuardEA} standardizes identifiers and data types via hierarchical abstraction to bolster Type-2 detection, while CCDive~\cite{10908670} employs a transformation localization mechanism to sustain high precision under heavy code refactoring. Complementarily, CCStokener~\cite{WANG2023111618} constructs semantic tokens by fusing local structure with n-gram modeling, specifically targeting improved recall for Medium Type-3 (MT3) clones.
To mitigate the inherent insensitivity of neural models to logic-preserving structural changes, alternative approaches utilize syntactic and control-flow properties as "logical skeletons." Tamer~\cite{Hu2023FineGrainedCC} introduces block-level Abstract Syntax Tree (AST) segmentation for subtree matching, whereas StoneDetector~\cite{10473623} encodes control-flow paths using Dominator Trees to ensure robustness against intricate syntactic variations. Notwithstanding these advancements in explicit structural grounding, such methodologies often struggle with Type-4 semantic clones—where functional equivalence persists despite radical structural divergence—due to their limited capacity for deep semantic intent modeling. Bridging the divide between rigid structural constraints and flexible semantic representations remains a pivotal challenge for advancing fine-grained clone classification.
\subsection{Code Representation Learning and Semantic Analysis}
\label{subsec:2_4}
Pre-trained models (PTMs) have significantly advanced the capture of functional equivalence. While CodeBERT~\cite{feng-etal-2020-codebert} pioneered bi-modal representation learning, subsequent frameworks like GraphCodeBERT~\cite{guographcodebert} and UniXcoder~\cite{guo-etal-2022-unixcoder} integrated data-flow and syntactic structures to sharpen sensitivity toward complex semantic variants. Conversely, Toma~\cite{10.1145/3597503.3639114} demonstrated the robustness of lexical priors as discriminative anchors in low-to-medium similarity scenarios.
Nevertheless, lexical-centric approaches remain ill-equipped for Type-4 clones characterized by radical structural divergence. Furthermore, the prevailing focus on binary classification leaves the efficacy of PTMs across the nuanced similarity gradients of BigCloneBench largely under-explored. This necessitates an ensemble paradigm capable of harmonizing lexical-based boundary calibration with semantic-based functional recall. Orchestrating robust arbitration strategies to resolve inter-feature conflicts thus represents a critical frontier for fine-grained clone detection.
\subsection{Hybrid Representations and Multi-modal Learning}
\label{subsec:2_5}
Integrating multi-dimensional signals into hybrid representations has become essential for addressing the expressivity limits of unimodal models in semantic clone identification. Pioneer frameworks like FA-AST~\cite{9054857} augmented AST topologies with control- and data-flow edges using graph neural networks, while FCCA~\cite{9146274} employed attention mechanisms for dynamic multi-feature fusion. More recently, DSFM~\cite{10.1145/3597503.3639215} pivoted toward fine-grained interaction modeling via AST subtree decomposition and Factorization Machines. Despite these advancements, neural discriminators often struggle with "deterministic calibration" at decision margins, failing to ground high-uncertainty predictions in programmatic "physical facts" or heuristic priors. Harmonizing deep semantic embeddings with deterministic heuristics to ensure both semantic richness and structural rigor thus remains a pivotal challenge for fine-grained clone detection.
\subsection{LLM-driven Code Clone Detection}
\label{subsec:2_6}
The reasoning prowess of Large Language Models (LLMs) is shifting clone detection from static representation learning toward inference-based discrimination. Evaluations \cite{10576803} suggest that while GPT-4 excels in functional comprehension, its instability in extreme Type-4 cases positions it as a high-level adjudicator rather than a general-purpose filter. Despite advancements via generative augmentation \cite{10336319} and instruction tuning \cite{liang2025hyclonebridgingllmunderstanding}, LLMs still rely on "semantic intuition," complicating alignment with the rigid thresholds of BigCloneBench. Furthermore, existing hybrid systems often utilize static triggers rather than adaptive, confidence-based arbitration.
\par In summary, bridging the chasm between "semantic hallucination" and "programmatic constraints" remains a formidable challenge. Neural fusion strategies frequently suffer from discriminative erosion at ambiguous boundaries due to a lack of deterministic anchors to resolve feature conflicts.
\par We address these gaps with a multi-modal fusion and intelligent arbitration framework. By integrating ML-based priors, AST skeletons, and CodeBERT embeddings, we architect a system with dynamic confidence sensing. Central to this approach is a DeepSeek-driven arbitration mechanism that performs logical rectification on low-confidence instances, achieving high-precision identification across the entire fine-grained clone spectrum.
\section{Methodology}
\label{sec:part3}
\subsection{Task Definition and Rationale for Fine-grained 7-class Classification}
\label{subsec:3_1}
This work centers on fine-grained 7-class code clone detection, an objective that necessitates the precise categorization of code pairs into Non-clone, Type-1, Type-2, and four distinct similarity gradients of Type-3/4 \cite{7332459}. While such granularity is indispensable for tracking sophisticated code evolution, the inherent overlap in feature spaces—particularly between MT3 and WT3—poses significant discriminative challenges.
To address these complexities, we propose a composite framework that integrates "semantic-heuristic-structural" multi-dimensional representations with a logical arbitration layer. Each dimension provides complementary strengths:
\begin{itemize}
\item \textbf{Deep Semantic Representations (CodeBERT):} By leveraging pre-trained weights to capture functional equivalence \cite{feng-etal-2020-codebert}, CodeBERT establishes a robust semantic baseline. Its bidirectional attention mechanism strikes an optimal balance between expressivity and throughput, offering a more stable backbone for high-frequency comparisons than purely generative models.
\item \textbf{ML-based Heuristic Priors:} This component utilizes statistical features to establish deterministic anchors based on literal facts, effectively mitigating the over-generalization tendencies of neural models in high-similarity scenarios.
\item \textbf{AST-based Syntactic Skeletons:} By extracting hierarchical syntax, this dimension ensures resilient logical alignment against identifier renaming and minor structural refactoring.
\end{itemize}

Despite multi-modal integration, neural discriminators often falter in "gray zones" where feature signals conflict. To reconcile such decision uncertainty, we introduce the DeepSeek Large Language Model (LLM) as a Cognitive Arbitrator. Unlike the feature-centric nature of the primary components, the arbitrator performs high-order logical reasoning to rectify conflicting evidence when the primary model exhibits low confidence. This paradigm—coupling three-dimensional feature anchoring with adaptive logical arbitration (Fig.~\ref{fig:all})-is designed to fundamentally resolve discriminative degradation in fine-grained clone detection.

\begin{figure*}[t]
\centering
\includegraphics[width=1\textwidth]{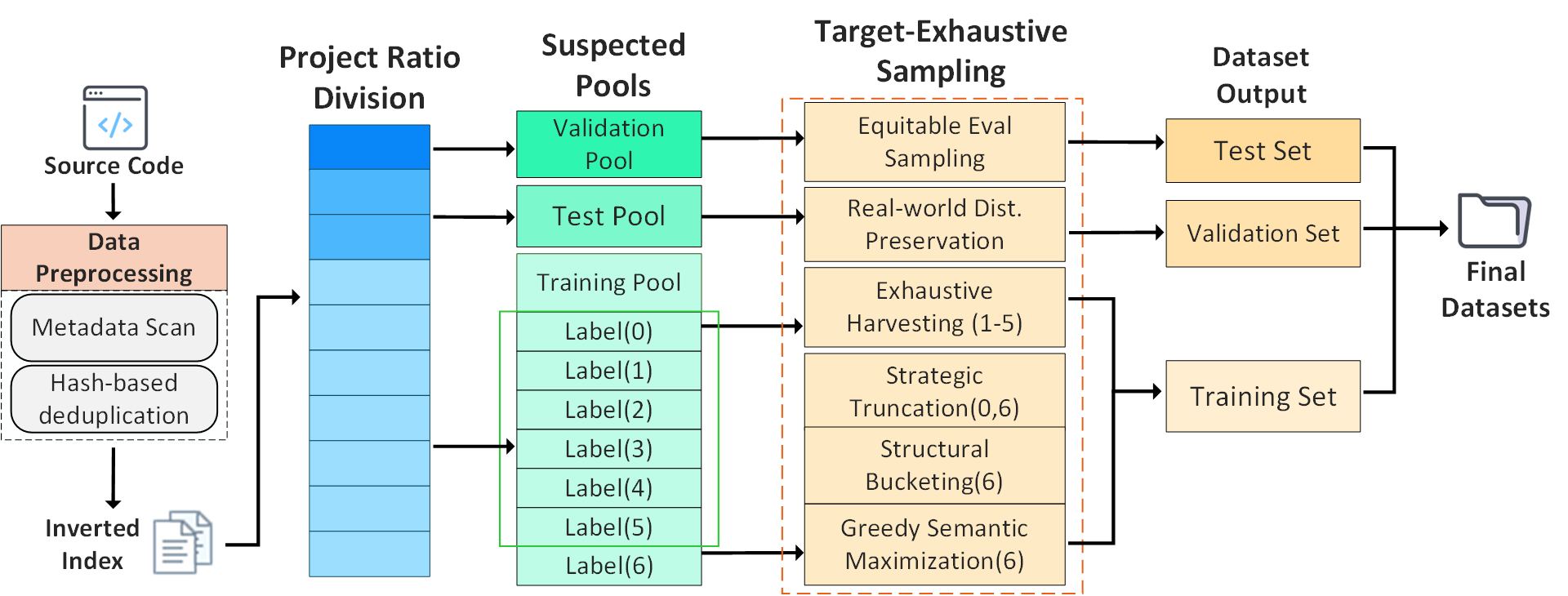}
\caption{Data Processing Pipeline of the Proposed Multi‑Dimensional Clone Detection Framework}
\vspace{0.3em} 
\footnotesize 
\textbf{Note:} Dist. = distribution.
\label{fig:pipeline}
\end{figure*}
\subsection{Dataset Construction and Preprocessing}
\label{subsec:data_prep}
Our empirical study is anchored in \textit{BigCloneBench} (BCB) \cite{7332459}, the de facto standard for evaluating code clone detection. To ensure empirical rigor and representative sampling, we developed a systematic preprocessing pipeline (illustrated in Fig.~\ref{fig:pipeline}) comprising four critical stages:
\subsubsection{Data Cleansing and Metadata Indexing}
We initially construct an inverted metadata index to facilitate efficient data scheduling. This is followed by a normalization phase where snippets under 200 characters are filtered to maintain semantic density. Finally, \texttt{SHA-256} hashing is employed for global deduplication to eliminate redundancy within the experimental corpus.
\subsubsection{Project-level Isolation Protocol}
To mitigate "coding-style leakage" and ensure an unbiased evaluation, we enforce a strict project-level isolation protocol across 4,781 independent software projects. These projects are partitioned into disjoint training (70\%), validation (10\%), and testing (20\%) pools. This strategy ensures that all test instances remain entirely unseen during the training phase, providing an objective measure of cross-project generalization.
\subsubsection{Goal-oriented Stratified Sampling}
We employ differentiated sampling strategies to balance learning efficiency with real-world distributional authenticity:
\begin{itemize}
\item \textbf{Training Set:} To alleviate class skewness, we implement strategic caps on majority classes (Label 0 and 6) at 40,000 and 25,000 pairs, respectively, while performing exhaustive collection for rare categories (Labels 1--5).
\item \textbf{Validation Set:} Balanced stratified sampling is applied with a target weight of 1,428 samples per label. For categories with insufficient data, the entire available population is retained.
\item \textbf{Testing Set:} To simulate large-scale retrieval in industrial scenarios, all compliant samples within the testing pool are preserved in their entirety without truncation or resampling.
\end{itemize}
\subsubsection{Diversity Optimization for Type-4 Semantic Clones}
Recognizing the inherent difficulty of Type-4 clone detection, we introduce structural binning and a greedy diversity optimization strategy:
\begin{itemize}
\item \textbf{Structural Binning:} Samples are partitioned into bins based on code length and complexity to ensure robust coverage across varying structural scales, from granular fragments to extensive functions.
\item \textbf{Maximizing Semantic Diversity:} Within each bin, we calculate Jaccard similarity \cite{Jaccard1912THEDO} and apply a greedy selection process that prioritizes samples with the lowest similarity to the currently selected set. This maximizes the semantic coverage of the training data, enhancing the model's resilience to long-tail semantic patterns.
\end{itemize}

The final distribution of the processed dataset is summarized in Table~\ref{tab:dataset_distribution}.
\begin{table}[ht]
\centering
\caption{Sample Distribution of the Experimental Dataset}
\label{tab:dataset_distribution}
\resizebox{0.95\linewidth}{!}{%
\begin{tabular}{l l r r r}
\toprule
Clone Type & Training & Validation & Testing (Real-world) \\
\midrule
Non-clone & 40,000 & 1,428 & 46,416 \\
Type-1 & 680 & 92 & 200 \\
Type-2 & 459 & 87 & 128 \\
Very Strong Type-3 & 936 & 99 & 264 \\
Strong Type-3 & 1,903 & 506 & 963 \\
Moderate Type-3 & 4,692 & 1,428 & 5,448 \\
Weak Type-3 + Type-4 & 25,000 & 1,428 & 881,258 \\
\midrule
\textbf{Total} & \textbf{73,670} & \textbf{5,068} & \textbf{934,677} \\
\bottomrule
\end{tabular}%
}
\end{table}

\subsection{Heterogeneous Feature Space Extraction}
\label{subsec:3_2}

\subsubsection{Deep Semantic Representation Layer (CodeBERT)}
\label{subsubsec:3_2_1}
We employ CodeBERT to derive deep semantic representations of code pairs. For each input function sequence, we extract the hidden vector $\mathbf{h}_{sem}\in\mathbb{R}^d$ from the final Transformer block's [CLS] token. This vector encodes high-dimensional semantic similarity and serves as the initial input for the 7-class classification task.

\subsubsection{Heuristic Lexical Priors}
\label{subsubsec:3_2_2}
To mitigate similarity bias in neural encoders, we introduce a 7-dimensional heuristic prior vector $\mathbf{s}\in\mathbb{R}^7$, obtained via LightGBM \cite{ke2017lightgbm} integration. It captures key lexical metrics, including token overlap and identifier similarity, providing a stable reference to distinguish ambiguous categories such as MT3 and WT3.

\subsubsection{Structural Statistics for Semantic Calibration}
\label{subsubsec:3_2_3}
We extract a 6-dimensional structural vector $\mathbf{v}_{ast}$ from code ASTs, comprising metrics like logical density, maximum depth, and total node count differences. Identifier-agnostic, these features calibrate the logical biases of semantic encoders. For VST3 (Class 3), AST stability ensures high recall for minor logical variants. Detailed extraction procedures and dimensional definitions are summarized in Table~\ref{tab:feature_summary}.
\begin{table*}[t]
\centering
\caption{Overview of Multi-dimensional Features}
\label{tab:feature_summary}
\resizebox{0.9\textwidth}{!}{%
\begin{tabular}{@{}>{\centering\arraybackslash}m{4cm} >{\centering\arraybackslash}m{10cm}@{}}
\toprule
\textbf{Feature Type} & \textbf{Key Features} \\
\midrule
Statistical Features (ML-based) & 
\begin{itemize}
    \item Code-level similarity: Jaccard, Dice, Overlap, Cosine, Levenshtein, TF-IDF Cosine
    \item Token statistics: unique tokens, total tokens, shared tokens
    \item Derived features: mean, standard deviation, max/min, token ratio/difference, structural interaction (cosine*(1-levenshtein))
\end{itemize} \\
Structural Features (AST-based) & 
\begin{itemize}
    \item Tree edit distance, number of common subtrees / subtree set similarity
    \item Node type statistics, node sequences / path sets
    \item Tree depth, width, number of leaves, structural hash fingerprints
\end{itemize} \\
Semantic Features (CodeBERT) & 
\begin{itemize}
    \item \text{[CLS]} vector, token-level average/max pooling
    \item Token sequence, intermediate hidden layer outputs
    \item Attention weight matrices
\end{itemize} \\
\bottomrule
\end{tabular}%
}
\end{table*}


\subsection{Heuristic-Guided Semantic Modulation}
\label{subsec:3_3}
To rectify potential semantic drift inherent in CodeBERT during fine-grained classification, we implement a modulation module based on Feature-wise Linear Modulation (FiLM) \cite{perez2018film}. This module utilizes the 7-dimensional heuristic prior vector $\mathbf{s}$ 
 to generate affine control signals, facilitating a dimension-wise recalibration of the deep semantic representation $\mathbf{h}_{sem}$:
\begin{IEEEeqnarray}{rCl}
\gamma, \beta & = & \text{MLP}_{\text{prior}}(\mathbf{s}) \\
\mathbf{h}_{sem}' & = & (1 + \gamma) \odot \mathbf{h}_{sem} + \beta
\end{IEEEeqnarray}
Within our 7-class detection framework, the FiLM layer functions as a categorical regulator. When heuristic priors indicate low lexical overlap, the generated $\gamma$
parameters suppress dimensions associated with strong clone signals, effectively anchoring the model's judgment in lexical grounding.

\subsection{Structurally-Aligned Fusion}
\label{subsec:3_4}
Building on the modulated semantic representations, we adopt a Structural Cross-Attention mechanism \cite{vaswani2017attention}, where the calibrated semantic vector $\mathbf{h}^{\prime}{sem}$ serves as the Query ($Q$), and the AST structural vector $\mathbf{v}{ast}$ forms the Key ($K$) and Value ($V$):
\begin{align} 
\label{eq:attention_calc} 
\text{Attn}(Q, K, V) &= \text{Softmax} \left( \frac{Q(\mathbf{h}'_{sem}) K(\mathbf{v}_{ast})^T}{\sqrt{d}} \right) V(\mathbf{v}_{ast}) \\ \label{eq:attention_fusion} \mathbf{h}_{fusion} &= \text{LayerNorm}(\mathbf{h}'_{sem} + \text{Attn}(Q, K, V)) \end{align}

This design ensures fine-grained classification integrates both semantic functional equivalence and structural consistency. By dynamically weighting attention, the model filters out samples that appear semantically similar but differ logically, achieving precise alignment across the spectrum from Label 3 to Label 6.






\subsection{Inference and Confidence-aware LLM Arbitration}
\label{subsec:3_5}
The fused representation $\mathbf{h}_{fusion}$ is fed into a 7-way Softmax layer to produce the predicted probability distribution 
\[
P = [p_0, p_1, \dots, p_6] \in \mathbb{R}^7
\] . To handle ambiguous samples residing near decision boundaries, we introduce a confidence-aware LLM arbitration mechanism triggered when 
$\max(P) < \tau$:
\begin{enumerate}
\item \textbf{Context Construction:} A diagnostic prompt is synthesized, incorporating the raw source code, the primary model's Top-3 candidate labels, and their respective probabilities.
\item \textbf{LLM Reasoning:} A Large Language Model (e.g., DeepSeek) performs high-order logical reasoning over the conflicting probabilistic evidence to finalize the classification.
\end{enumerate}

We employ DeepSeek-V3.2 \cite{deepseekai2025deepseekv32pushingfrontieropen} as the arbitration core. This model exhibits exceptional proficiency in long-sequence code understanding and logical deduction, while its Mixture-of-Experts (MoE) architecture ensures an optimal balance between reasoning accuracy and computational cost. Leveraging zero-shot reasoning capabilities, DeepSeek resolves evidential conflicts between multi-modal features, ensuring high-fidelity determination for boundary cases. This hierarchical approach—where the hybrid model acts as an efficient filter and the LLM serves as a precision arbitrator—strikes a critical balance between overall system throughput and detection accuracy.
\begin{table*}[!ht]
\centering
\caption{Fine-grained Per-Class Performance on BigCloneBench}
\label{tab:per_class_performance}
\begin{tabular}{lllcccc}
\toprule
\textbf{Category (Label)} & \textbf{Clone Type} & \textbf{Metric} & \textbf{Baseline (CodeBERT)} & \textbf{Our Model (Full)} & \textbf{Rel. Improv.} & \textbf{95\% CI (F1)} \\
\midrule
Label 0 & Non-clone             & P / R / F1 & 0.998 / 1.000 / 0.999 & 0.997 / 1.000 / 0.998 & -0.10\% & [0.995, 1.001] \\
Label 1 & Type-1 (Exact)        & P / R / F1 & 0.797 / 0.980 / 0.878 & 0.995 / 0.991 / 0.993 & 11.50\% & [0.991, 0.995] \\
Label 2 & Type-2 (Renamed)      & P / R / F1 & 0.690 / 0.768 / 0.726 & 0.824 / 0.855 / 0.839 & 11.40\% & [0.776, 0.903] \\
Label 3 & Very Strong T3        & P / R / F1 & 0.488 / 0.596 / 0.533 & 0.817 / 0.822 / 0.819 & 28.60\% & [0.807, 0.831] \\
Label 4 & Strongly T3           & P / R / F1 & 0.424 / 0.801 / 0.550 & 0.843 / 0.864 / 0.853 & 30.30\% & [0.825, 0.881] \\
Label 5 & Moderately T3         & P / R / F1 & 0.115 / 0.844 / 0.201 & 0.473 / 0.937 / 0.627 & 42.60\% & [0.591, 0.664] \\
Label 6 & Type-4 (Semantic)     & P / R / F1 & 1.000 / 0.952 / 0.975 & 1.000 / 0.992 / 0.996 & 2.10\%  & [0.995, 0.997] \\

\midrule
Overall & Macro Average         & F1         & 0.695 $\pm$ 0.023      & 0.875 $\pm$ 0.008       & 18.00\% & [0.865, 0.885] \\
\bottomrule
\end{tabular}
\end{table*}
\subsection{Evaluation Metrics}
To assess the multi-modal fusion framework and adaptive arbitration mechanism, we utilize a tiered suite of quantitative metrics:
\begin{itemize}
\item \textbf{Macro-F1 Score:} As the primary benchmark, the Macro-F1 provides an unweighted average across all seven categories—including structurally ambiguous types like MT3, WT3, and Type-4—to ensure a balanced assessment under class imbalance.
\item \textbf{Per-class Precision, Recall, and F1:} These granular metrics characterize the model’s discriminative fidelity in distinguishing subtle syntactic and semantic variations.
\item \textbf{95\% Confidence Intervals (CI) and Standard Deviation:} Predictive stability is quantified across multiple experimental runs with independent random seeds to assess aleatoric uncertainty, particularly within ill-defined clone categories.
\item \textbf{Top-$k$ Coverage:} This evaluates whether the ground truth resides within the top $k$ candidates. It serves as a vital indicator for the feasibility of our adaptive arbitration, ensuring that uncertain instances are refined by the LLM within a pruned hypothesis space and a constrained computational budget.
\end{itemize}

\subsection{Implementation Details}
Our deep fusion framework is implemented using PyTorch~\cite{pytorch} and Transformers~\cite{huggingface_transformers}. Experiments were conducted on a high-performance cluster with a 48-core CPU and four NVIDIA RTX 3090 GPUs (24GB VRAM each). To ensure robustness and reproducibility, each experiment was repeated with five random seeds, and results were averaged as the final metric.

The model was optimized with AdamW~\cite{Loshchilov2017DecoupledWD} ($\epsilon=1\times10^{-8}$) using a differential learning rate schedule: the CodeBERT encoder was fine-tuned at $2\times10^{-5}$, while fusion and classification layers used $1\times10^{-4}$. Training spanned 5 epochs with a global batch size of 32 and a linear scheduler incorporating 80 warmup steps. Automatic mixed-precision (AMP) was employed to improve memory efficiency and training throughput.

Additionally, we applied 0.1 label smoothing~\cite{7780677} to enhance prediction stability and mitigate overfitting, particularly along the nuanced decision boundaries of fine-grained clone classification.
\section{Experimental Results and Analysis}
This section presents the empirical findings to address the defined research questions.
\subsection{RQ1:How effective is the proposed multi-modal fusion framework, and what is the contribution of each component to the overall performance?}
The evaluation for RQ1—focusing on overall efficacy and the synergy among constituent modules—is structured in two stages. First, a system-level comparison between our full model and established baselines quantifies the magnitude of performance gains. Second, an ablation study deconstructs the impact of each heterogeneous component to reveal the underlying drivers and mechanisms of the observed improvements.



\subsubsection{RQ1.1: Performance Improvement over the Baseline}

We evaluate the multi-modal fusion framework from two perspectives: overall performance gain and the contribution of individual components.

On the 7-class task, we compared the \textbf{Full Model} against the \textbf{CodeBERT} baseline. Experiments were repeated across five random seeds, reporting both mean and variance to ensure robustness. As shown in Table~\ref{tab:per_class_performance}, the \textbf{Full Model} improved the Macro F1 from 0.695 to 0.875, an absolute gain of \textbf{18.0\%} with strong statistical significance ($p < 0.01$). The most substantial improvements occurred in medium-difficulty structural categories (Labels 2--4), with F1 gains of 11.4\%, 28.6\%, and 30.3\%, respectively.
\begin{figure*}[t]
\centering
\begin{subfigure}{0.49\linewidth}
    \centering
    \includegraphics[width=\linewidth]{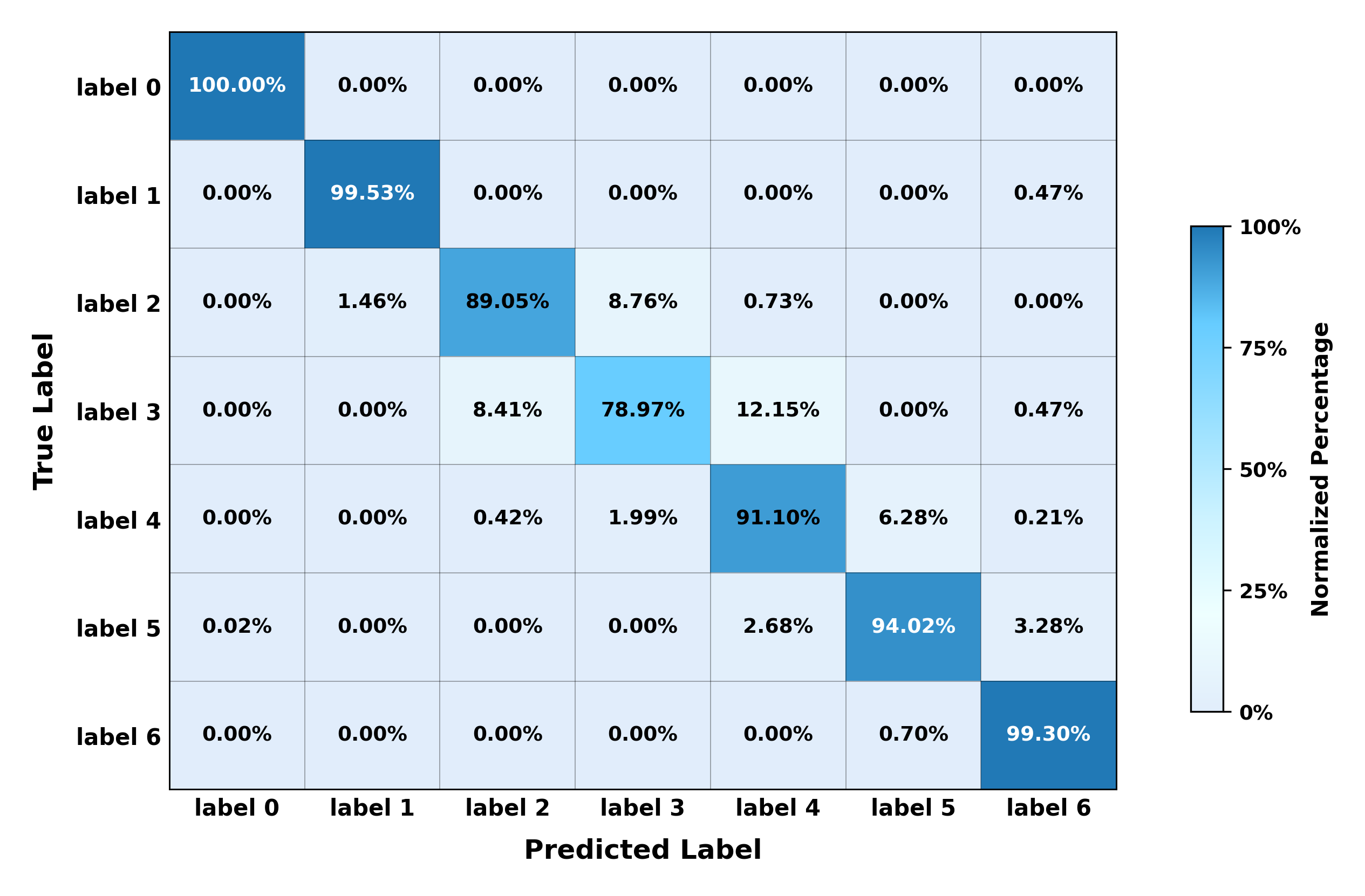}
    \caption{Normalized confusion matrix of the proposed model}
    \label{fig:norm_cm}
\end{subfigure}
\hfill
\begin{subfigure}{0.49\linewidth}
    \centering
    \includegraphics[width=\linewidth]{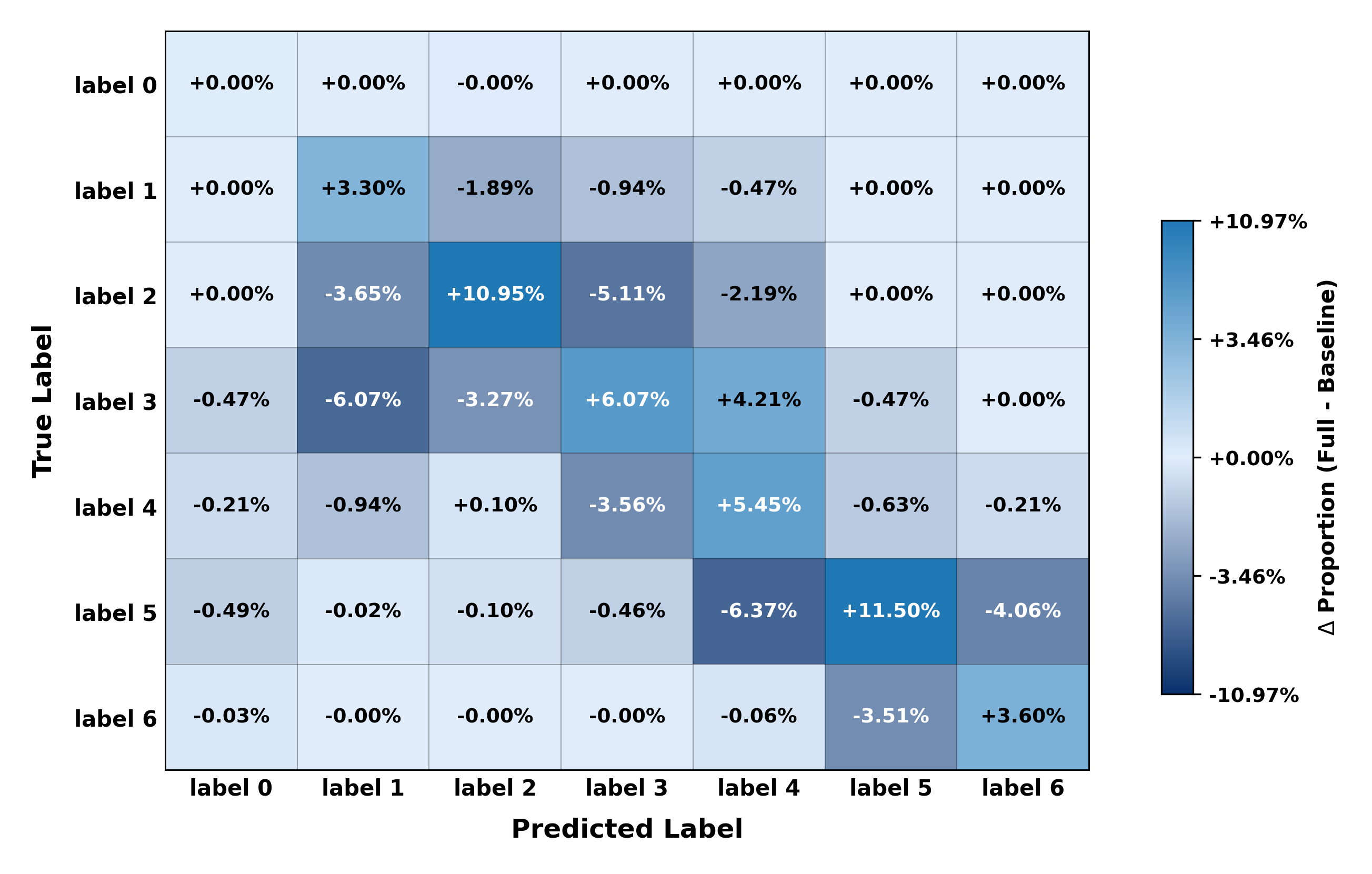}
    \caption{Performance difference compared with the baseline}
    \label{fig:diff_cm}
\end{subfigure}
\caption{Confusion matrix analysis on the test set. 
(a) Normalized confusion matrix of the proposed model.
(b) Difference matrix illustrating performance improvements over the baseline.}
\label{fig:confusion_analysis}
\end{figure*}

Experimental results demonstrate that our model achieves robust and substantial gains across all categories. As shown in the normalized confusion matrix (Figure~\ref{fig:norm_cm}), diagonal accuracies exceed \textbf{89\%} for every category. Specifically, Non-Clone (Label 0), Type-1 (Label 1), and Type-4 (Label 6) reached \textbf{100\%}, \textbf{99.53\%}, and \textbf{99.30\%}, respectively. Even for intermediate categories (Labels 2--5), the model maintains strong discriminative power, achieving 89.05\%, 78.97\%, 91.10\%, and 94.02\% accuracy. Misclassifications are mostly confined to structurally adjacent labels, highlighting sensitivity to fine-grained distinctions. The minor 0.1\% F1 drop for Label 0 is negligible compared to the overall improvements.

The difference confusion matrix (Figure~\ref{fig:diff_cm}) clarifies the sources of these gains. Labels 3, 4, and 5 show diagonal improvements of \textbf{+6.07\%}, \textbf{+5.45\%}, and \textbf{+11.50\%}, accompanied by a marked reduction in false positives. For instance, the misclassification of Label 5 as Label 4 decreased by 6.37\%, directly contributing to the Macro-F1 increase.

Moreover, the \textbf{Full Model} exhibits enhanced stability over the baseline, with the standard deviation narrowing from $\pm0.023$ to $\pm0.008$ and the 95\% confidence interval converging to \textbf{[0.865, 0.885]}. This indicates that multi-modal feature fusion, serving as a decision calibration mechanism, not only elevates the performance ceiling but also mitigates inference volatility in complex samples, enhancing \textbf{discriminative stability}. For the most challenging Type-4 semantic clones (Label 6), the model attains an F1 of \textbf{0.996}, recall of \textbf{99.2\%}, and perfect precision of \textbf{1.000}. Incorporating structural constraints preserves the model’s capacity to capture deep functional intent while reinforcing classification boundaries via multi-dimensional feature alignment. Overall, the framework overcomes baseline limitations in complex clone scenarios while retaining high-fidelity detection for simpler types.

\begin{table*}[t]
\centering
\caption{Per-Clone-Type F1 Comparison Across Model Variants}
\label{tab:model_variants_perclass}
\begin{tabular}{lccccccccc}
\toprule
\textbf{Model Variant} & \textbf{Macro-F1} & \textbf{$\Delta$} & \textbf{Label 1} & \textbf{Label 2} & \textbf{Label 3} & \textbf{Label 4} & \textbf{Label 5} & \textbf{Label 6} & \textbf{Label 0} \\
\midrule
Our Model (Full)       & 0.875 & /       & 0.993 & 0.839 & 0.819 & 0.853 & 0.627 & 0.996 & 0.998 \\
AST-only               & 0.291 & -0.584  & 0.381 & 0.000 & 0.377 & 0.274 & 0.210 & 0.658 & 0.139 \\
ML-only        & 0.834 & -0.041  & 0.993 & 0.793 & 0.789 & 0.784 & 0.657 & 0.989 & 0.839 \\
M$_{\text{struct}}$ (w/o struct)               & 0.853 & -0.022  & 0.993 & 0.835 & 0.823 & 0.832 & 0.495 & 0.993 & 1.000 \\
M$_{\text{struct}}$ (w/o ML-heur)            & 0.807 & -0.068  & 0.925 & 0.810 & 0.721 & 0.811 & 0.396 & 0.990 & 0.998 \\
M$_{\text{struct}}$ (naive)     & 0.861 & -0.014  & 0.940 & 0.806 & 0.797 & 0.890 & 0.605 & 0.995 & 0.997 \\
CodeBERT               & 0.695 & -0.180  & 0.878 & 0.726 & 0.533 & 0.550 & 0.201 & 0.975 & 0.999 \\
\bottomrule
\end{tabular}
\end{table*}
\begin{table*}[t]
\centering
\caption{BigCloneBench 7-Class Clone Detection Overall Performance (Mean $\pm$ Std over 5 Seeds)}
\label{tab:bigclonebench_performance}
\begin{tabular}{lcccccc}
\toprule
\textbf{Method} & \textbf{Accuracy} & \textbf{Macro-Precision} & \textbf{Macro-Recall} & \textbf{Macro-F1} & \textbf{95\% CI (Macro-F1)} & \textbf{p-value} \\
\midrule
CodeBERT (Baseline)       & 0.953 $\pm$ 0.005  & 0.644 $\pm$ 0.024 & 0.849 $\pm$ 0.018 & 0.695 $\pm$ 0.023 & [0.666, 0.723] & / \\
AST-only                  & 0.509 $\pm$ 0.047  & 0.299 $\pm$ 0.043 & 0.432 $\pm$ 0.078 & 0.291 $\pm$ 0.070 & [0.204, 0.378] & 1.0 \\
ML-only                    & 0.994 $\pm$ 0.006  & 0.806 $\pm$ 0.009 & 0.875 $\pm$ 0.011 & 0.834 $\pm$ 0.006 & [0.827, 0.841] & 0.0 \\
M$_{\text{struct}}$ (w/o struct)      & 0.986 $\pm$ 0.005  & 0.823 $\pm$ 0.022 & 0.923 $\pm$ 0.009 & 0.853 $\pm$ 0.017 & [0.832, 0.874] & 0.0 \\
M$_{\text{struct}}$ (w/o ML-heur)     & 0.980 $\pm$ 0.002  & 0.771 $\pm$ 0.021 & 0.904 $\pm$ 0.016 & 0.807 $\pm$ 0.019 & [0.783, 0.831] & 0.0 \\
M$_{\text{struct}}$ (naive)           & 0.991 $\pm$ 0.002  & 0.831 $\pm$ 0.019 & 0.921 $\pm$ 0.008 & 0.861 $\pm$ 0.015 & [0.843, 0.880] & 0.0 \\
Our Model (Full)           & 0.992 $\pm$ 0.001  & 0.850 $\pm$ 0.009 & 0.923 $\pm$ 0.012 & 0.875 $\pm$ 0.008 & [0.865, 0.885] & 0.0 \\
\bottomrule
\end{tabular}
\end{table*}
\subsubsection{RQ1.2: Component Contribution Analysis via Ablation Study}
To dissect the functional roles of the individual components within our multi-modal framework, we conducted a systematic ablation study by incrementally removing key modules. The results indicate that different information sources perform distinct roles and collaboratively drive the overall performance gains in fine-grained code clone detection, as detailed in Tables~\ref{tab:model_variants_perclass} and \ref{tab:bigclonebench_performance}.

Primarily, the heuristic priors extracted via machine learning (e.g., statistical matrices such as Jaccard similarity) constitute the bedrock of the model's discriminative power. Removing this ML-prior module (M$_{\text{struct}}$ (w/o ML-heur)) leads to a substantial decline in Macro-F1 from \textbf{0.875} to \textbf{0.807} (a decrease of $-0.068$). This performance degradation is particularly acute in categories with high classification complexity, such as VST3 (Label 3) and MT3 (Label 5), where F1 scores plummeted by \textbf{0.102} and \textbf{0.231}, respectively. This observation underscores that heuristic statistical features provide indispensable discriminative anchors, effectively bolstering the model's capacity to recognize intricate code variants.

Furthermore, structural information derived from Abstract Syntax Trees (ASTs) serves as a critical stabilizer. While AST features alone are insufficient for complex detection tasks (yielding an F1 of only \textbf{0.291}), their integration as a structural branch provides a performance lift of approximately \textbf{0.022} Macro-F1 over the structure-less variant (M$_{\text{struct}}$ (w/o struct)). Notably, in the MT3 (Label 5) category—where decision boundaries are most ambiguous—the AST branch contributes a significant gain of \textbf{0.132}. This confirms that structural features capture topological invariants, thereby compensating for the perceptual deficiencies of pure semantic models when confronted with radical syntactic transformations.

Analyzing across the clone gradient reveals a distinct complementary relationship between heterogeneous features. In high-certainty categories (Labels 0, 1, and 6), the semantic backbone provides nearly perfect discrimination (F1$\geq$\textbf{0.993}), where structural and heuristic features remain largely "non-intrusive." However, in structural variant regions of medium complexity (Labels 2–4), semantic and structural signals couple to mitigate ambiguities induced by identifier renaming and structural refactoring. For the most challenging MT3 (Label 5) category, multi-modal fusion substantially enhances classification reliability through cross-modal alignment.

A comparison of fusion strategies highlights that even a naive concatenation model ($M_{naive}$), despite utilizing all features, underperforms our full model with a Macro-F1 of \textbf{0.861}. This suggests that in high-dimensional feature spaces, the absence of an effective interaction mechanism can lead to "signal drowning," where dominant signals obscure fine-grained functional cues. By contrast, our framework employs FiLM-based modulation and synergistic alignment to achieve dynamic calibration, ensuring that heterogeneous information sources form an orthogonal and complementary relationship.
Lastly, multi-modal fusion significantly enhances prediction consistency from a statistical perspective. The \textbf{95\% Confidence Interval (CI)} for the CodeBERT baseline across multiple random seeds spans a width of \textbf{0.057} ($[0.666, 0.723]$), reflecting high uncertainty in boundary cases. In contrast, the full model's CI narrows to \textbf{0.020} ($[0.865, 0.885]$), achieving high statistical significance ($p < 0.001$). These findings demonstrate that by incorporating multi-dimensional constraints and adaptive calibration, our framework not only elevates the performance ceiling but also substantially bolsters the stability of the model's decision-making process.
\subsection{RQ2:How does the confidence-aware adaptive arbitration mechanism resolve discriminative conflicts?}
\label{subsec:rq2}

Building on the empirical results above, we observe that multi-modal fusion substantially improves detection, yet residual aleatoric uncertainty remains in classes with highly ambiguous structural boundaries. To address this, we study the cost–benefit trade-off of an adaptive arbitration mechanism driven by a large language model (DeepSeek). To rigorously evaluate how the arbitration layer balances performance gains against computational overhead, our analysis concentrates on two aspects:

\begin{itemize}
\item {Quantifying the primary model's uncertainty and deterministically configuring the arbitration trigger threshold; }
\item {Measuring the resulting shifts in global and per-class performance—with particular focus on Label 5—after applying the chosen threshold.}
\end{itemize}

\begin{table*}[htb]
\centering
\small
\renewcommand{\arraystretch}{1.05}
\caption{Validation set results (per-class). Confidence and confidence intervals are shown as percentages. Top-3 coverage and classification metrics (Precision / Recall / F1) are reported per-class.}
\label{tab:validation-results}
\begin{tabular}{lrrrrrrrr}
\toprule
Label & Mean Conf & CI Low & CI High & Top-3 & Precision & Recall & F1 & Support \\
 & (\%) & (\%) & (\%) & Coverage (\%) & (\%) & (\%) & (\%) &  \\
\midrule
Label 0 & 91.499\% & 91.496\% & 91.502\% & 100.000\% & 100.000\% & 100.000\% & 100.000\% & 1428 \\
Label 1 & 92.219\% & 90.565\% & 93.874\% & 99.091\% & 100.000\% & 99.091\% & 99.543\% & 110 \\
Label 2 & 72.226\% & 65.736\% & 78.715\% & 100.000\% & 85.294\% & 78.378\% & 81.690\% & 74 \\
Label 3 & 70.175\% & 64.161\% & 76.188\% & 100.000\% & 79.245\% & 80.000\% & 79.621\% & 105 \\
Label 4 & 83.507\% & 81.797\% & 85.217\% & 99.650\% & 89.358\% & 92.645\% & 90.972\% & 571 \\
Label 5 & 87.371\% & 86.520\% & 88.222\% & 100.000\% & 97.262\% & 94.538\% & 95.881\% & 1428 \\
Label 6 & 90.368\% & 89.938\% & 90.798\% & 100.000\% & 97.729\% & 99.440\% & 98.577\% & 1428 \\
\bottomrule
\end{tabular}

\end{table*}


\subsubsection{RQ2.1:Determination of the Arbitration Trigger Strategy based on Model Confidence}
Uncertainty Distribution and Stability Clustering. Guided by the validation statistics in Table~\ref{tab:validation-results}, we quantitatively characterized the primary model’s uncertainty via predicted confidence levels and their 95\% confidence intervals (CIs). Our observations reveal that Labels 0, 1, and 6 constitute a high-stability cluster with mean confidence exceeding 90\% and near-optimal F1 scores. Given their remarkably narrow CIs, these categories (comprising 57.7\% of the dataset) are handled sufficiently by the primary encoder, precluding the necessity for further arbitration.
Discriminative Bottlenecks and Candidate Identification. Conversely, Labels 2–5 emerge as the discriminative bottleneck of the model. Specifically, Labels 2 and 3 exhibit substantial aleatoric uncertainty, with mean confidence hovering around 70\% and CI volatility ranges exceeding 12 percentage points. Although Labels 4 and 5 show signs of stabilization, this high-uncertainty cluster—representing 42.3\% of the population—remains the primary candidate for LLM-based adjudication.

Hypothesis Space Pruning Rationale. Crucially, the global weighted Top-3 coverage reaches 99.94\%, ensuring that the ground truth almost certainly resides within the top three candidates even when the primary prediction is erroneous. By pruning the hypothesis space into a highly condensed candidate set, the primary model effectively reduces the reasoning complexity and noise susceptibility for the subsequent LLM-based arbitrator, facilitating a more robust and precise rectification process.
Based on these empirical insights, we propose a Validation-Driven Hybrid Trigger Strategy. The system circumvents arbitration for high-confidence categories (L0/1/6) while selectively invoking the DeepSeek arbitrator for samples characterized by low confidence or dispersed CIs (primarily L2--5). This targeted allocation of computational resources ensures that logical rectification is focused on instances with the highest discriminative ambiguity, thereby optimizing the trade-off between global inference throughput and detection fidelity.

To evaluate the efficacy and robustness of the proposed arbitration strategies, we developed two standardized prompt archetypes (Listing~\ref{lst:unified-arbitration-prompt}). The first archetype incorporates the primary model’s probabilistic priors—specifically candidate labels and their respective confidence scores—to facilitate evidence-based adjudication by DeepSeek. In contrast, the second archetype relies solely on raw source code for independent classification, serving as a baseline for ablation studies and control group analysis. Both designs enforce a rigid JSON schema for the output, facilitating seamless automated parsing while ensuring the reproducibility of our empirical findings.
\definecolor{lightgray}{gray}{0.95}
\begin{lstlisting}[
caption={Unified arbitration prompt supporting DeepSeek and LLM-only modes},
label={lst:unified-arbitration-prompt},
basicstyle=\ttfamily\small,
breaklines=true,
backgroundcolor=\color{lightgray},
xleftmargin=2mm,
xrightmargin=2mm,
showspaces=false,
showtabs=false,
tabsize=2,
columns=fixed
]
prompt = f"""
you are an expert arbitrator in program analysis. You will determine the clone type (0-6) 
of the following code pair. Your task can follow two modes:
DeepSeek mode:
Combine code semantics with a high-performance model's prior probabilities
to produce a final decision (0-6) with reasoning and confidence.
LLM-only mode:
Use ONLY the code content, ignoring any model predictions or probabilities.
--- Code 1 ---
{code1_short}
--- Code 2 ---
{code2_short}
--- Optional Model Prior Probabilities (for DeepSeek mode) ---
{ref_info}
--- BigCloneBench Labels (0-6) ---
[1] Type-1: identical syntax except whitespace/comments
[2] Type-2: identical syntax except identifiers/literals
[3] VST3: syntactic similarity [90%,100%)
[4] ST3: syntactic similarity [70%,90%)
[5] MT3: syntactic similarity [50%,70%)
[6] WT3 / Type-4: semantic clone (<50% similarity)
[0] Non-Clone: completely different logic
Return STRICT JSON only:
{
  "mode": "<DeepSeek or LLM-only>",
  "thought": "<analysis of core code differences>",
  "prediction": <0-6 integer>,
  "confidence": <0.0-1.0>,
  "explanation": "<key reasoning>",
  "probabilities": [p0,p1,p2,p3,p4,p5,p6]
}
"""
\end{lstlisting}
\begin{figure*}[t]
\centering
\begin{subfigure}{0.49\linewidth}
    \centering
    \includegraphics[width=\linewidth]{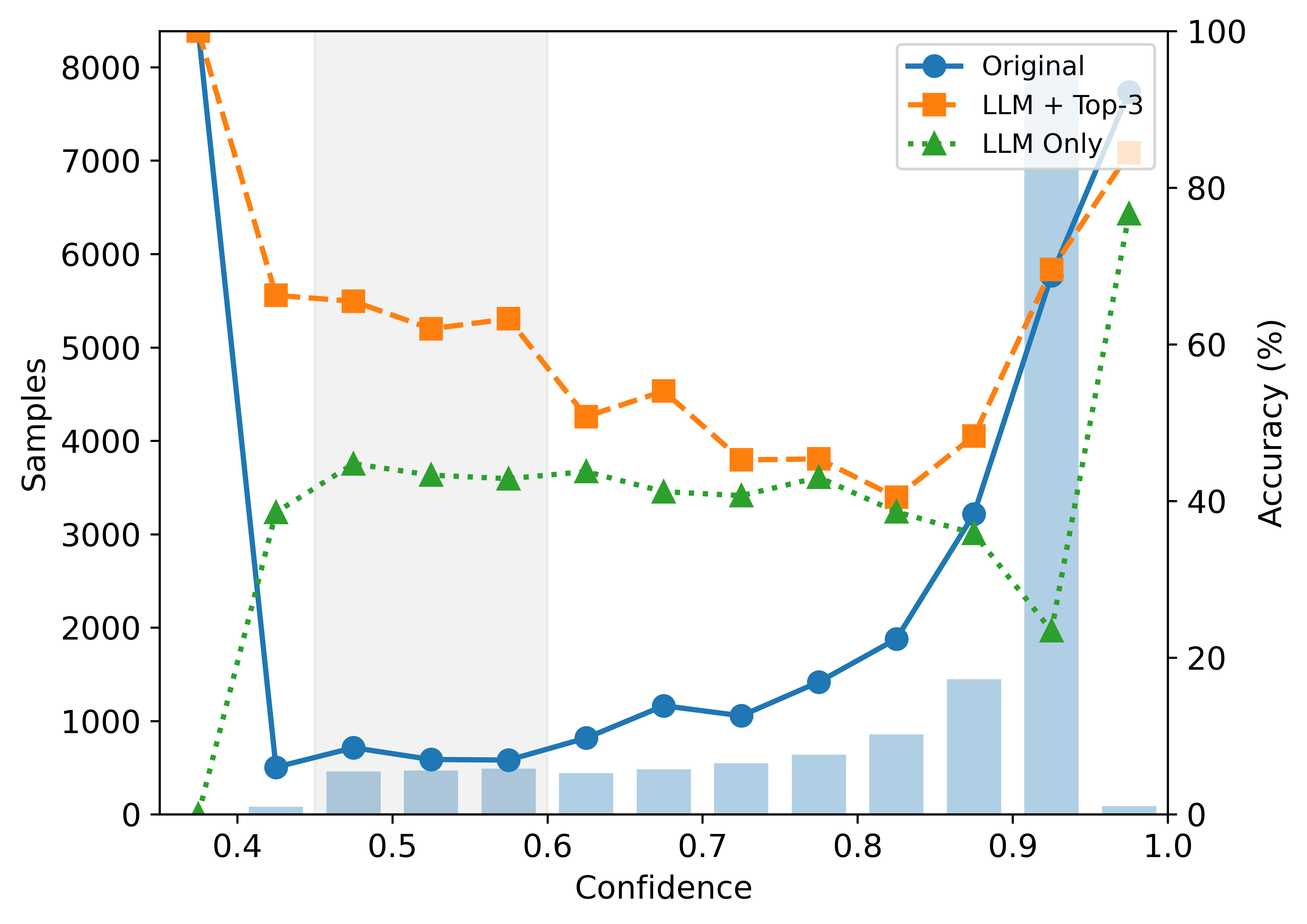}
    \caption{Accuracy across confidence bins}
    \label{fig:confidence-acc}
\end{subfigure}
\hfill
\begin{subfigure}{0.49\linewidth}
    \centering
    \includegraphics[width=\linewidth]{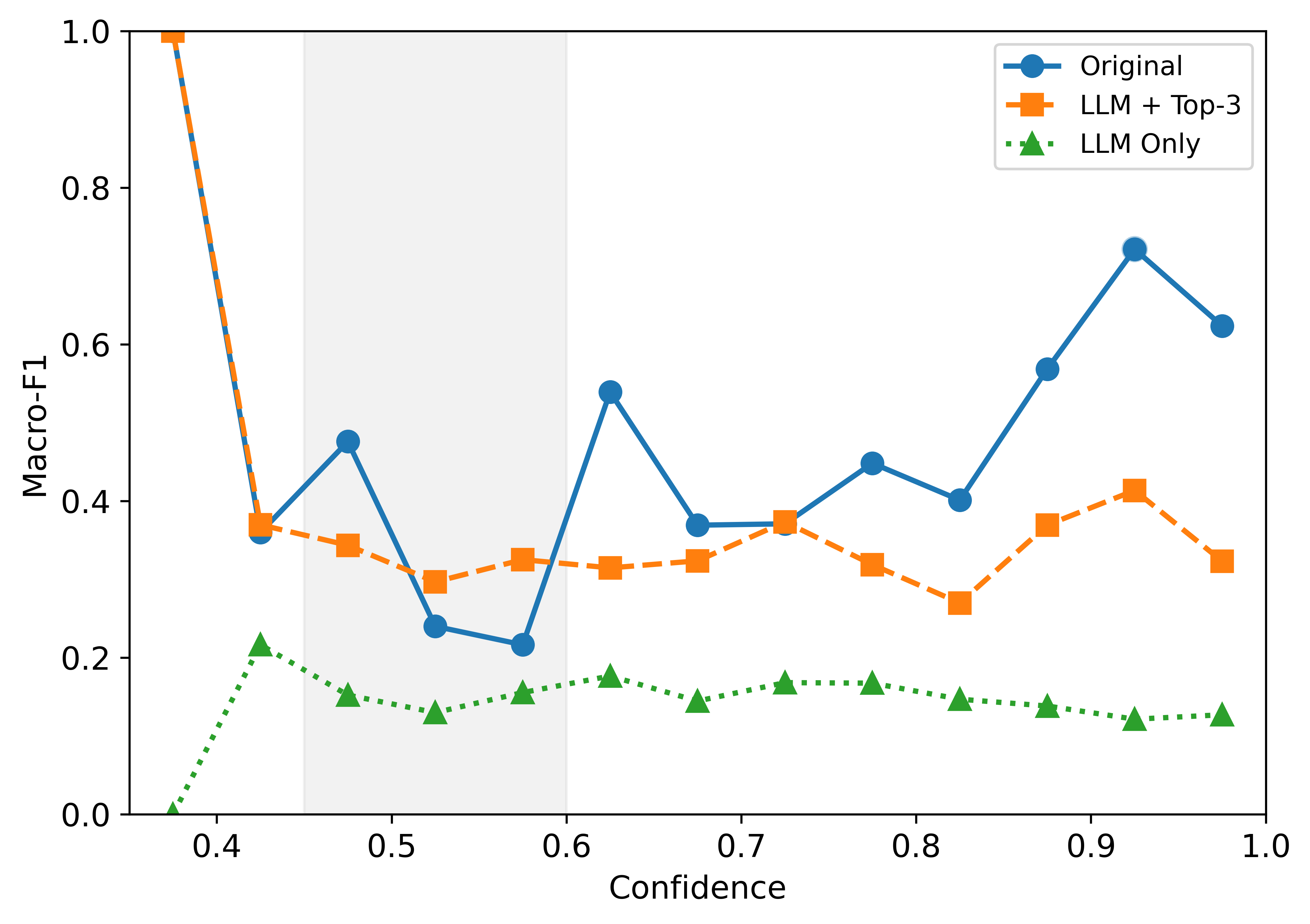}
    \caption{Macro-F1 across confidence bins}
    \label{fig:confidence-f1}
\end{subfigure}
\caption{Performance behavior across confidence intervals on the validation set. 
The left figure shows prediction accuracy across bins, while the right figure 
illustrates the Macro-F1 variation used to determine the arbitration threshold.}
\label{fig:confidence-analysis}
\end{figure*}

\subsubsection{\textit{RQ2.2: Maximizing Performance via Adaptive Arbitration}}
\begin{figure}[t]
\centering
\begin{subfigure}{0.45\linewidth}
    \centering
    \includegraphics[width=\linewidth]{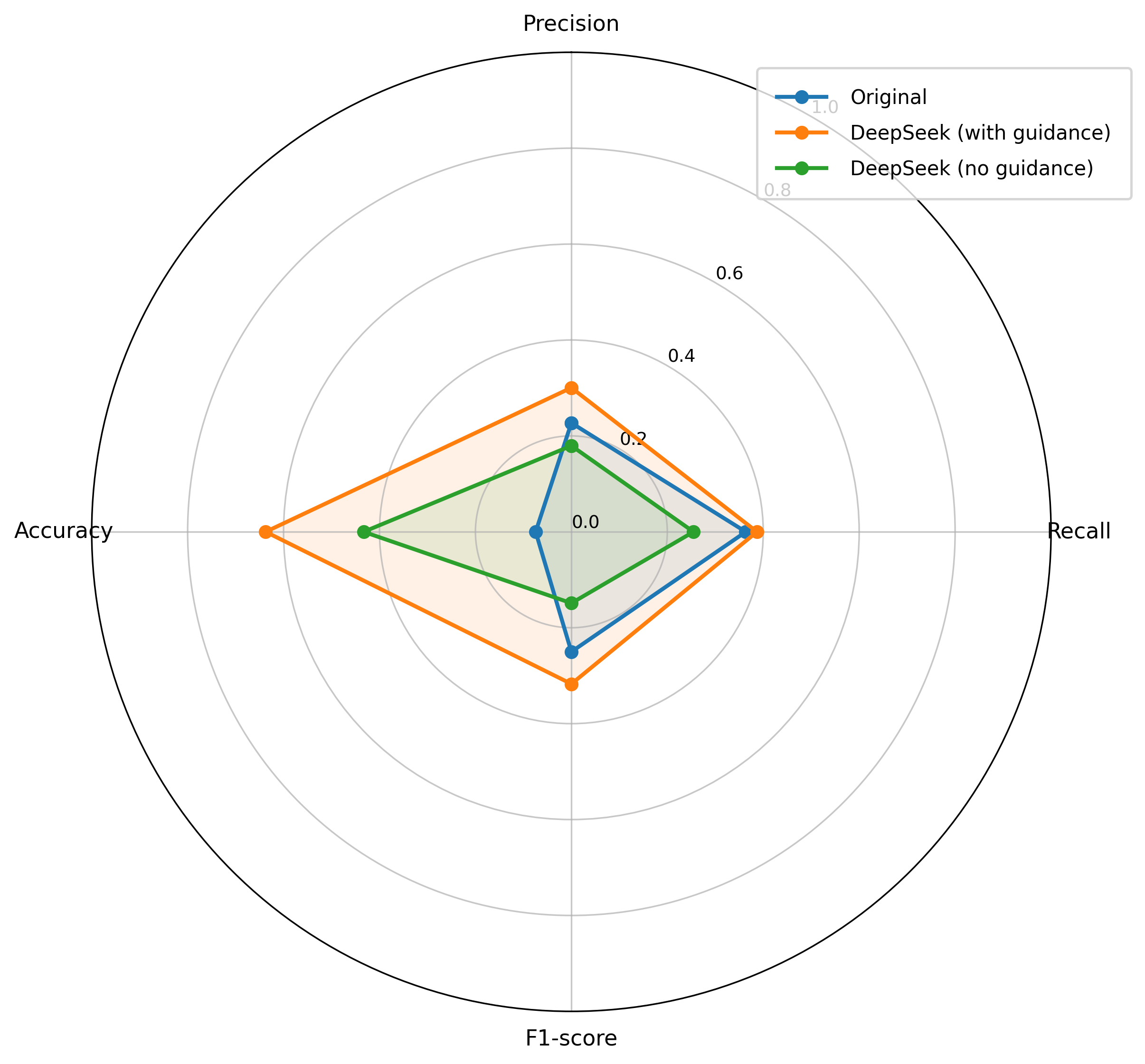}
    \caption{Macro‑averaged Metrics \& Overall Accuracy}
    \label{fig:radar_maccro}
\end{subfigure}
\hfill
\begin{subfigure}{0.45\linewidth}
    \centering
    \includegraphics[width=\linewidth]{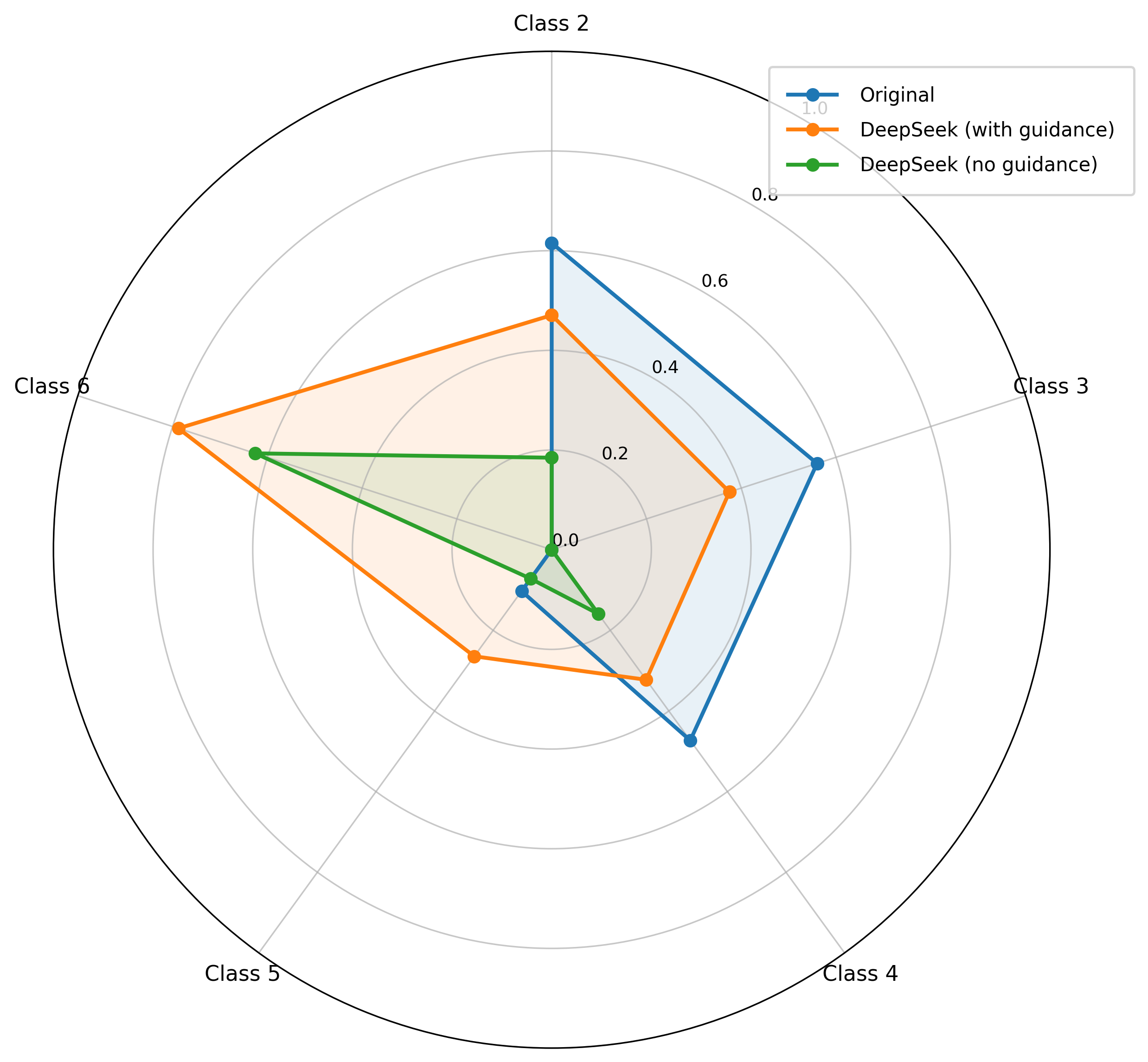}
    \caption{Per‑class F1‑score Robustness}
    \label{fig:radar_preclass}
\end{subfigure}
\caption{Performance behavior across confidence intervals on the validation set. 
The left figure shows prediction accuracy across bins, while the right figure 
illustrates the Macro-F1 variation used to determine the arbitration threshold.}
\label{fig:radar-analysis}
\end{figure}

To investigate the efficacy of adaptive arbitration, we stratified the test set by prediction confidence and compared three paradigms: the standalone primary model, blind LLM re-classification, and LLM+Top-3 guided arbitration. Empirical evidence indicates that samples with confidence scores below 0.6 (comprising 0.16\% of the test set) exhibit a precipitous decline in predictive fidelity, with the primary model yielding a meager Weighted Macro-F1 of \textbf{0.311}. A performance inflection point is observed at the \textbf{0.6} threshold (Fig.~\ref{fig:confidence-analysis}), beyond which the marginal utility of LLM intervention diminishes. Consequently, we designated 0.6 as the optimal trigger threshold, striking a pragmatic balance between performance augmentation and computational economy by targeting only the 0.15\% highest-entropy samples.
Effectiveness Comparison on the Low-Confidence Subset
Within the cohort where confidence$<$0.6, the ground-truth distribution is highly skewed, with Label 6 (Type-4) accounting for 89.0\%.Detailed performance metrics and distribution are shown in Figure~\ref{fig:radar-analysis}.
\begin{itemize}
\item Standalone Model Limitations: Predictions are erroneously confined to Labels 2-5, resulting in a deficient \textbf{7.45\%} accuracy. Specifically, Label 5 suffers from a critically low precision (0.055), leading to excessive false positives.
\item Blind DeepSeek Arbitration: Re-classification without prior guidance improves accuracy to \textbf{43.31\%} but introduces a severe bias toward the majority class, causing the Macro-F1 for minority categories to collapse to 0.149.
\item Guided DeepSeek Arbitration: Integrating probabilistic priors from the primary model enables DeepSeek to achieve a superior discriminative equilibrium (Accuracy: \textbf{63.74\%}, Weighted F1: 0.734). Notable improvements include a three-fold increase in Label 5 precision and the preservation of sensitivity toward rare patterns (Labels 2-4), which outperformed the blind baseline by nearly 3x.
\end{itemize}
\begin{table*}[t]
\centering
\caption{Performance Comparison: Original vs Replacement Strategies}
\label{tab:performance_comparison}
\begin{tabular}{lccccc}
\toprule
\textbf{Metric} & \textbf{Original} & \textbf{Replace 2,3,4,5} & $\Delta$ 2,3,4,5 & \textbf{Replace 5 only} & $\Delta$ 5 only \\
\midrule
Accuracy           & 0.9920 & 0.9929 & +0.0009 & 0.9929 & +0.0009 \\
Macro Precision    & 0.8521 & 0.8470 & -0.0051 & 0.8543 & +0.0022 \\
Macro Recall       & 0.9357 & 0.9335 & -0.0022 & 0.9354 & -0.0003 \\
Macro F1           & 0.8803 & 0.8792 & -0.0011 & 0.8837 & +0.0034 \\
Weighted Precision & 0.9960 & 0.9960 & +0.0000 & 0.9960 & +0.0000 \\
Weighted Recall    & 0.9920 & 0.9929 & +0.0009 & 0.9929 & +0.0009 \\
Weighted F1        & 0.9934 & 0.9940 & +0.0006 & 0.9940 & +0.0006 \\
\bottomrule
\end{tabular}
\end{table*}
Industrial-Scale Validation
Evaluations on the full test set (934,677 samples) demonstrate that selectively arbitrating only the low-confidence Label 5 instances improves global Macro-F1 by 0.0034. Conversely, broad-spectrum arbitration of all low-confidence samples leads to a decline in Macro-Precision. This underscores the necessity of precision-targeted intervention—focusing exclusively on the most challenging categories—to maintain global performance integrity without incurring prohibitive costs (Table~\ref{tab:performance_comparison}).

As evidenced by our experimental evaluation on the test set, the proposed confidence-aware adaptive arbitration mechanism—leveraging model priors to guide DeepSeek—substantially bolsters the model’s discriminative fidelity on low-confidence bottleneck samples. Concurrently, this approach maintains a negligible computational footprint, necessitating high-order LLM invocation for less than 0.2\% of the total instances. By facilitating an optimal equilibrium between detection accuracy, inference overhead, and categorical fairness, this strategy offers a viable and high-precision paradigm for cost-efficient code clone detection in large-scale industrial scenarios.

\section{Threats to Validity}
Several factors may limit the generalizability of our findings. External validity is constrained by the Java-centric nature of BigCloneBench and the scarcity of high-fidelity fine-grained datasets for cross-language validation. Construct validity concerns arise from ambiguous Type-3/4 boundaries across benchmarks, which may introduce categorical bias. Regarding internal validity, our selective DeepSeek arbitration yielded modest absolute metric gains; however, this hierarchical paradigm successfully optimizes the trade-off between detection fidelity and computational overhead—a critical requirement for cost-sensitive industrial deployment.
\section{Discussion}
Empirical evidence suggests that the efficacy of fine-grained clone detection hinges on the strategic orchestration of heterogeneous signals. By positioning the LLM as an "on-demand" adjudicator, our framework maintains high-throughput inference while leveraging superior reasoning to resolve decision-boundary ambiguities. Notably, invoking probability-guided arbitration for a mere 0.2\% of high-entropy samples yields a strategic gain of over 0.3\% in global Macro-F1. This utility is highly category-specific: while marginal for Type-2/3 clones where discriminative models suffice, the arbitrator is vital for Type-4 semantic clones, where it reconciles radical structural divergence with functional equivalence. Given the prevalence of binary classification in existing literature, this study underscores that the selective orchestration of discriminative and generative paradigms is a robust trajectory for tackling complex semantic clone detection
\section{Conclusion}
This study introduces a multi-dimensional fusion framework integrating ML-based heuristic priors, AST structures, and CodeBERT semantic embeddings to tackle the challenges of 7-class fine-grained clone detection. Empirical results demonstrate that this multi-modal orchestration effectively compensates for unimodal deficiencies, propelling the Macro-F1 score from 0.695 to 0.875, with particularly pronounced gains in the complex MT3 category. A pivotal innovation is the confidence-triggered DeepSeek arbitration mechanism; by leveraging the primary model's output probability distribution to guide logical rectification, this informed approach significantly outperforms baseline arbitration strategies that lack probabilistic grounding. This collaborative paradigm reconciles high-throughput discriminative encoding with the sophisticated reasoning of generative models. Future work will refine uncertainty-based triggering strategies and extend the framework to cross-lingual contexts and large-scale code retrieval.
\bibliographystyle{IEEEtran}
\bibliography{reference}

\end{CJK}  
\end{document}